\documentclass[prd,preprint, superscriptaddress,floatfix,eqsecnum,nofootinbib,12pt]{revtex4}
\usepackage{hyperref}
\usepackage{amssymb,amsmath,amsthm,graphicx,ulem}
\pdfoutput=1
\usepackage{graphicx,subfigure}
\usepackage{ulem}
\usepackage{epsfig}
\usepackage{amsmath}
\usepackage{amsfonts}
\usepackage{amssymb}
\usepackage[usenames]{color}
\usepackage[letterpaper,left=2cm,right=2cm,top=2.32cm,bottom=2.32cm]{geometry}
\usepackage[T1]{fontenc}

\newcommand{\beq}{\begin{equation}}
\newcommand{\eeq}{\end{equation}}
\newcommand{\be}{\begin{equation}}
\newcommand{\ee}{\end{equation}}
\newcommand{\beqa}{\begin{eqnarray}}
\newcommand{\eeqa}{\end{eqnarray}}
\newcommand{\beqar}{\begin{eqnarray*}}
\newcommand{\eeqar}{\end{eqnarray*}}
\newcommand{\bea}{\begin{eqnarray}}
\newcommand{\eea}{\end{eqnarray}}

\newcommand{\dd}{\textrm{d}}

\newcommand{\nn}\nonumber

\begin{document}

\title{The Real No-Boundary Wave Function \\ in Lorentzian Quantum Cosmology}

\author{J. Diaz Dorronsoro}
\email{juan.diaz@fys.kuleuven.be}
\affiliation{Institute for Theoretical Physics, KU Leuven, 3001 Leuven, Belgium}
\author{J. J. Halliwell}
\email{j.halliwell@imperial.ac.uk}
\affiliation{Blackett Laboratory, Imperial College, London SW7 2BZ, UK}
\author{J. B. Hartle}
\email{hartle@physics.ucsb.edu}
\affiliation{Department of Physics, UCSB, Santa Barbara, CA 93106, USA}
\author{T. Hertog}
\email{thomas.hertog@fys.kuleuven.be}
\affiliation{Institute for Theoretical Physics, KU Leuven, 3001 Leuven, Belgium}
\author{O. Janssen}
\email{opj202@nyu.edu}
\affiliation{Center for Cosmology and Particle Physics, NYU, NY 10003, USA}

\begin{abstract}
\noindent 
It is shown that the standard no-boundary wave function has a natural expression in terms of a Lorentzian path integral with its contour defined by Picard-Lefschetz theory. The wave function is real, satisfies the Wheeler-DeWitt equation and predicts an ensemble of asymptotically classical, inflationary universes with nearly-Gaussian fluctuations and with a smooth semiclassical origin.
\end{abstract}

\maketitle

\section{Introduction} \label{introsec}
The conventional formulation of the no-boundary wave function \cite{HH1983} involves a Euclidean path integral taken along an appropriate complex contour, whose choice is constrained by certain reasonable physical considerations \cite{HarHal1990}. It would be appealing to have an expression of the no-boundary wave function that is based on a Lorentzian path integral \cite{Brown1990,Halliwell1990}. This would yield a new route towards a more precise formulation of the wave function, perhaps using a holographic approach in which the dual is most naturally defined on the future boundary of spacetime, well into the asymptotically classical, Lorentzian domain of superspace \cite{Maldacena2002,Anninos2012, Hertog2011,Hartle:2012tv,McFadden2009}.

In this paper we evaluate the no-boundary wave function in a homogeneous isotropic minisuperspace model consisting of gravity coupled to a positive cosmological constant and scalar field matter. Our starting point is the Lorentzian path integral
\begin{equation} \label{amplitude}
\Psi  = \int_\mathcal{C} \mathcal{D}g \mathcal{D}\phi ~ e^{i S[g, \phi ]/\hbar} ~,
\end{equation}
with appropriate boundary conditions on the geometries $g$ and matter fields $\phi$ and taken along an appropriate contour ${\cal C}$ which ensures its convergence. When properly constructed, the path integral \eqref{amplitude} generates solutions to the Wheeler-DeWitt equation \cite{Halliwell:1990qr}. For the specific case of minisuperspace models with configuration space coordinates $q$ and $\phi$ that we consider here, the path integral is conveniently written in terms of an integral over the lapse function $N$ \cite{Teitelboim:1981ua,Teitelboim:1983fh,Teitelboim:1983fi,HalliwellWdW1988},
\begin{equation}
\Psi = \int_\mathcal{C}\dd N \int \mathcal{D}q \mathcal{D}\phi ~ e^{i S[N,q, \phi ]/\hbar} ~,
\end{equation}
where the functional integral over $q$ and $\phi$ has the form of a standard non-relativistic path integral between fixed initial and final data and fixed time interval $N$. 

We consider a Lorentzian contour for the lapse integral that runs along the entire real axis and avoids the singularity at $N=0$ by going below this point. Using Picard-Lefschetz theory to rigorously evaluate the path integral in the saddle point approximation we show this yields a solution of the Wheeler-DeWitt equation. Specifically, with no-boundary conditions this yields the usual, real no-boundary wave function describing two identical copies of an ensemble of asymptotically classical, inflationary universes \cite{Hawking1983,HHH2008} with Gaussian fluctuations \cite{Halliwell:1984eu}.

Our method resembles closely that of recent work of Feldbrugge et al. \cite{FLT1,FLT2} who also used Picard-Lefschetz theory in a Lorentzian framework but took the contour over $N$ to be half-infinite. This leads to a significantly different result, which we will discuss.
We also briefly comment on the broader implications of a Lorentzian viewpoint on the no-boundary wave function.

\section{de Sitter minisuperspace model} \label{dSMSPsec}
We first consider a homogeneous isotropic minisuperspace approximation to gravity coupled to a positive cosmological constant $\Lambda$ and no matter fields. Following \cite{Halliwell1988} we write the metric of the minisuperspace model as
\begin{equation} \label{metricansatz}
\dd s^2 = -\frac{N(\tau)^2}{q(\tau)}\dd \tau^2 + q(\tau) \dd \Omega_3^2 ~,
\end{equation}
where $\dd \Omega_3^2$ is the metric on the unit three-sphere. With this parametrization of the metric the Einstein-Hilbert action reads\footnote{We set $8 \pi G_N = 1$.}
\begin{equation}
S = 2 \pi^2 \int_0^1 \dd \tau ~ N \left(3 - \frac{3}{4N^2}\dot{q}^2 -  \Lambda q \right)~.
\end{equation}

Since the action is quadratic in the field, we may evaluate the path integral \eqref{amplitude} in this minisuperspace model exactly by parametrizing a general path as a deviation from the classical path satisfying the boundary conditions $q(0) = q_0, q(1) = q_1$. The resulting path integral over the deviations is that of a free particle which may be evaluated trivially. Imposing no-boundary initial conditions $q_0 = 0$, the integral over $q$ yields \cite{Halliwell1988}
\begin{equation}\label{wavefunction}
\Psi_{NB}^{\ }(q_1) = \sqrt{\frac{3\pi i}{2\hbar}}\int_\mathcal{C}\frac{\dd N}{\sqrt{N}}e^{2 \pi^2 i S_0(N,q_1)/\hbar}~,
\end{equation}
where the reduced action is
\begin{equation} \label{lapseint}
S_0 = \frac{\Lambda^2}{36}N^3+\left(3-\frac{\Lambda q_1}{2} \right) N - \frac{3 q_1^2}{4N} ~,
\end{equation}
and where the remaining integral over the lapse obviously depends on the contour $\mathcal{C}$. 

In the semiclassical limit one can evaluate the integral in \eqref{wavefunction} using the method of steepest descent. Picard-Lefschetz theory provides a rigorous way of determining which saddle points contribute in the semiclassical approximation for a given $\mathcal{C}$ \cite{Witten2010}. The method consists of identifying the curves of steepest ascent and steepest descent (a.k.a. Lefschetz thimbles) emanating from each of the (non-degenerate) saddle points, enabling one to determine how the original contour $\mathcal{C}$ should be deformed into a sum of steepest descent contours (which are lines of constant phase of the exponential part of the integrand). On each of these the familiar Gaussian approximation may then be applied in the small $\hbar$ limit. Whenever a steepest ascent curve intersects the integration contour $\mathcal{C}$ an odd number of times, the saddle point from which it emerges contributes to the integral.

\begin{figure}[htb]
\includegraphics[width=3.6in]{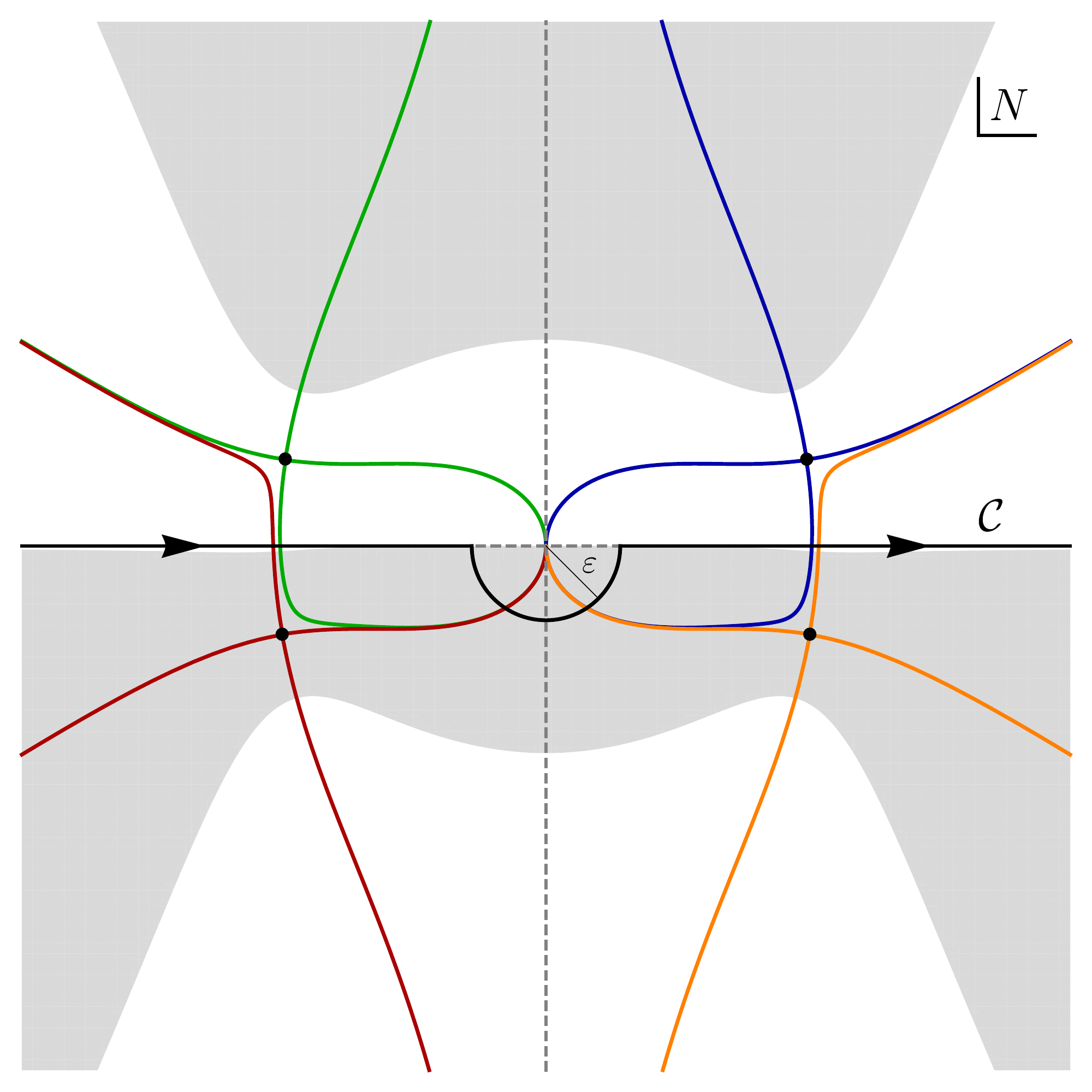} 
\caption{The four saddle points \eqref{dSMSPsaddles} of the contour integral \eqref{wavefunction} in the complex $N$-plane together with their steepest ascent and descent curves. In the shaded region $\text{Re}(i S_0) > 0$, suggesting divergent behaviour of an integral along a contour running to complex infinity or to the essential singularity at the origin in this domain. The Lorentzian contour $\mathcal{C}=(-\infty,+\infty)^{\downarrow}$ avoids the origin by passing along a parametrically small circle of radius $\varepsilon$ below that point. Analyticity away from $N=0$ ensures that the value of $\varepsilon>0$ does not affect the outcome of the integral. The parameter values $\Lambda = 3$ and $q_1 = 10$ were taken and to lift the degeneracy we considered the perturbation $S_0 \rightarrow S_0 + i \, .02 N^2$.}
\label{PL1}
\end{figure}

Motivated by physical considerations which will become clear below we take the integral \eqref{wavefunction} over the lapse along the contour $\mathcal{C}=(-\infty,+\infty)^{\downarrow}$ indicated in Figure \ref{PL1} by the black curve.\footnote{We take the branch cut of the square root function in \eqref{wavefunction} to lie along the positive imaginary $N$-axis.} For $q_1 > 3 / \Lambda$, which corresponds to the regime of superspace where we expect the wave function to predict classical evolution, the reduced action $S_0$ has four saddle points in the complex $N$-plane, located at
\begin{equation} \label{dSMSPsaddles}
N_s = \pm\frac{3}{\Lambda}\left(i\pm\sqrt{\frac{\Lambda q_1}{3} - 1}\right)~.
\end{equation}
The corresponding steepest ascent and descent curves of the integrand in \eqref{wavefunction} in the complex $N$-plane are also illustrated in Figure \ref{PL1}.

\begin{figure}[htb]
\includegraphics[width=4in]{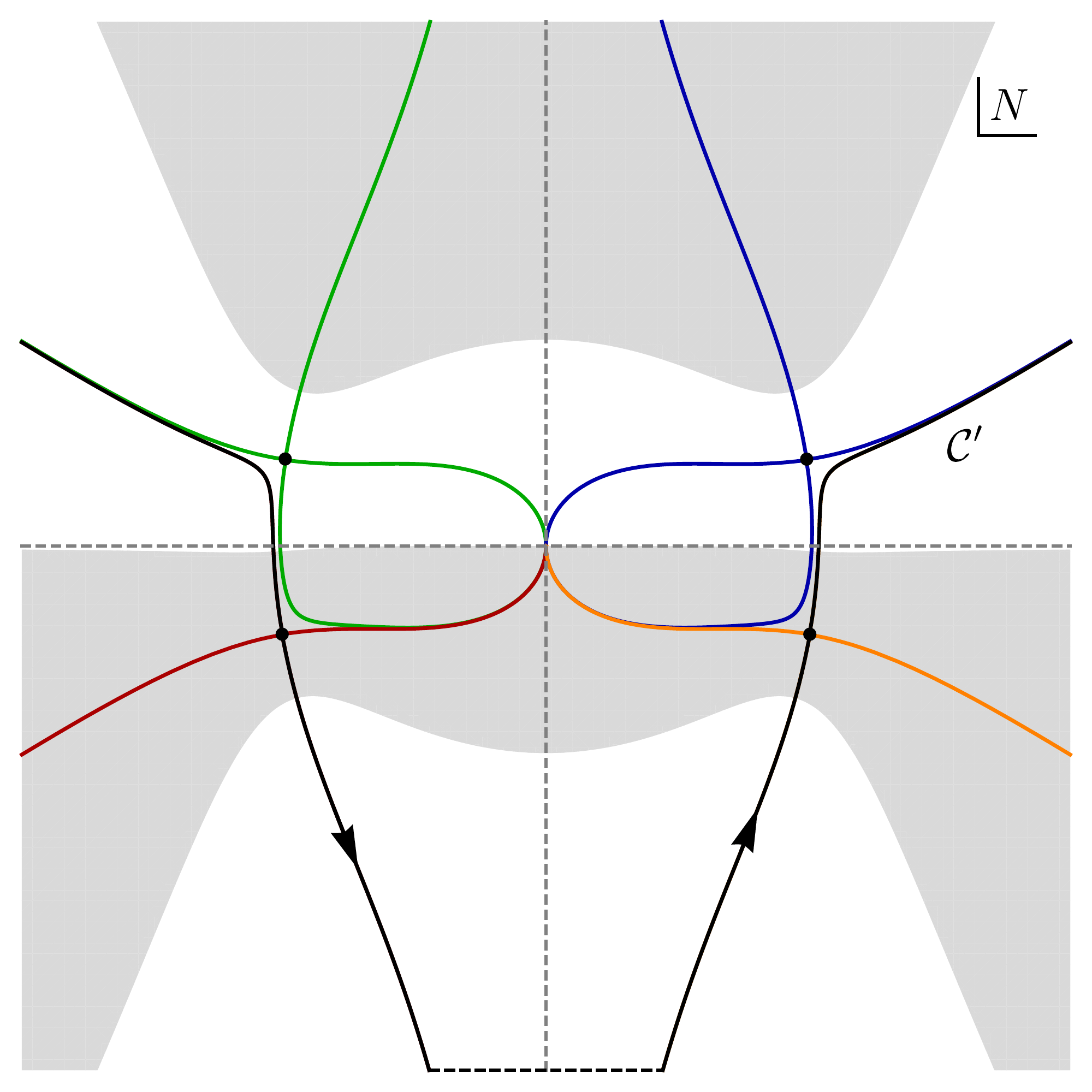} 
\caption{The continuous deformation ${\cal C}'$ implied by Picard-Lefschetz theory of the original contour $\mathcal{C}=(-\infty,+\infty)^{\downarrow}$ that passes through the two saddle points in the lower half complex $N$-plane. The two Lefschetz thimbles both tend to the negative imaginary axis at complex infinity, where $\text{Re}(i S_0) \rightarrow -\infty$. The contribution coming from the `arc at infinity' connecting the Lefschetz thimbles and the positive and negative real $N$-axes vanishes.}
\label{PL2}
\end{figure}

For a given contour $\mathcal{C}$ the Picard-Lefschetz prescription can be used to identify rigorously which saddle points contribute to the integral.
The continuous deformation ${\cal C}'$ of $\mathcal{C}$ implied by Picard-Lefschetz theory is shown in Figure \ref{PL2}.
The only subtlety in the Picard-Lefschetz analysis comes from the fact that the reduced action $S_0$ is a real function of $N$, $\overline{S_0(N)} = S_0(\bar{N})$, which leads to a degeneracy in the steepest ascent and descent curves \cite{Tanizaki2014}. This can be remedied by including a small symmetry breaking perturbation in the action and then take the limit in which this perturbation vanishes \cite{Witten2010,Serone2017}. Independently of how the perturbation is taken to zero, we find that the two saddle points in the lower half complex $N$-plane contribute.\footnote{As the perturbation tends to zero, the deformed contour also runs over the saddle points in the upper half plane. However contributions from these saddles are exponentially suppressed compared to those in the lower half plane, so we may safely ignore these \cite{dingle1973asymptotic}.}
This in fact corresponds to the obvious deformation suggested by examining Figure \ref{PL1}.  Semiclassically, therefore, and taking into account the angle at which the Lefschetz thimble passes through the saddle point and the prefactors coming from the Gaussian integrals, we find for the wave function
\begin{align} \label{HHsaddle}
\Psi_{NB}^{\ } (q_1) = \frac{e^{+12\pi^2/\hbar\Lambda}}{(\Lambda q_1 / 3 - 1)^{1/4}}\cos\left[ \frac{12 \pi^2}{\hbar \Lambda} \left(\frac{ \Lambda q_1}{3} - 1\right)^{3/2} + \frac{3 \pi}{4} \right] \left[ 1 + \mathcal{O}(\hbar) \right] ~.
\end{align}

This is real and precisely equal to the no-boundary wave function familiar from its Euclidean formulation.\footnote{In hindsight it follows from \cite{Halliwell1988} that an essentially Lorentzian contour yields the no-boundary wave function.} In particular, at sufficiently large values of the scale factor the wave function evaluated on a $q_1$-surface takes a WKB form and predicts classical scale factor evolution \cite{Hawking1983,HHH2008}. It serves as an initial condition for future and past evolution which in this very simple minisuperspace model is simply Lorentzian de Sitter space. The de Sitter universe comes with the familiar no-boundary weighting\footnote{One may wonder how the Picard-Lefschetz prescription can select exponentially enhanced saddle point contributions. Note however that our original contour passes just below the origin, where the real part of the exponent of the integrand behaves as $+1 / \varepsilon$. By a downwards flow from here we are led to the usual no-boundary saddle points, which are exponentially enhanced to leading order in $\hbar$. By contrast in the analysis of \cite{FLT1} the real part of the exponent of the integrand is zero, so one can only flow to exponentially suppressed saddles.} but since there is only one history in this model, this doesn't mean much. Below we therefore extend our model to include a scalar field. \newpage

Finally we verify that our choice of contour yields a solution to the Wheeler-DeWitt equation. The Wheeler-DeWitt equation for the de Sitter minisuperspace model is \cite{Halliwell1988,FLT1}
\begin{equation}
\hbar^2\frac{\partial^2 \Psi}{\partial q_1^2}+12\pi^4(\Lambda q_1-3)\Psi=0~.
\end{equation}
An explicit computation using \eqref{wavefunction} shows that
\begin{equation}\label{wdw}
\hbar^2\frac{\partial^2 \Psi}{\partial q_1^2}+12\pi^4(\Lambda q_1-3)\Psi=6\pi^2i\sqrt{\frac{3\pi i}{2}}\left[\frac{e^{2\pi^2 i S_0(N,q_1)/\hbar}}{\sqrt{N}}\right]_{-\infty}^{\infty}.
\end{equation}
Since the right-hand side vanishes, we see that our solution $\Psi$ indeed satisfies the Wheeler-DeWitt equation. If by contrast we had taken ${\cal C}$ to be a half-infinite contour, we would have found a delta function on the right-hand side in \eqref{wdw} producing a Green's function $G$ of the Wheeler-DeWitt equation rather than a genuine solution $\Psi$ \cite{FLT1}.

\section{Scalar Matter} \label{scalarmattersec}
We now generalize this calculation to include scalar field matter. As matter content we consider a single scalar field $\phi$, minimally coupled to gravity via the action
\begin{align}
	S = \frac{1}{2} \int \dd^4 x \sqrt{-g} \left( R - \left( \partial \phi \right)^2 - \, 2 V(\phi)\right) + \int \dd^3 y \sqrt{g^{(3)}} K ~.
\end{align}
We concentrate on the analytically tractable model of \cite{Garay1990} in which the potential takes the form
\begin{equation}
V(\phi)= \Lambda\cosh \left( \sqrt{\frac{2}{3}} \, \phi \right) ~.
\end{equation}
For small values of $\phi$ this matter model reduces to a cosmological constant $\Lambda$ and a massive scalar field with mass $m^2 = 2 H^2$, where $H^2 \equiv \Lambda/3$. A smart change of variables
\begin{align}
x&=\frac{\sqrt{3}}{2}~q \, \cosh\left( \sqrt{\frac{2}{3}} \, \phi \right)~,\label{x}\\
y&=\frac{\sqrt{3}}{2}~q \, \sinh\left( \sqrt{\frac{2}{3}} \, \phi \right)~,\label{y}
\end{align}
renders the action quadratic and allows one to perform the path integral over $x$ and $y$ explicitly \cite{Garay1990}. Imposing no-boundary initial conditions $q_0=0$ one finds
\begin{equation} \label{2fieldintegral}
\Psi_{NB}^{\ }(q_1,\phi_1) = \frac{2 \pi}{\hbar} \int_\mathcal{C}\frac{\dd N}{N}e^{2 \pi^2 i S_0(N,q_1,\phi_1)/\hbar}~,
\end{equation}
where 
\begin{equation}
S_0 = \frac{H^4}{4} N^3 - 3 \left[ \frac{H^2 q_1}{2} \cosh \left( \sqrt{\frac{2}{3}} \, \phi_1 \right) - 1 \right] N - \frac{3 q_1^2}{4N}~,
\end{equation}
and $\mathcal{C}=(-\infty,+\infty)^{\downarrow}$ is the same contour we considered in the de Sitter minisuperspace model of Section \ref{dSMSPsec}. The saddle points $N_s$ of this action satisfy a quartic equation and are given by
\begin{equation} \label{scalarsaddles}
N_s = \pm \frac{1}{H^2} \left( \sqrt{F-\lambda} \pm \sqrt{F+\lambda} \right) ~,
\end{equation}
where we have defined
\begin{equation}
F (q_1,\phi_1) \equiv \frac{H^2 q_1}{2} \cosh \left( \sqrt{\frac{2}{3}} \, \phi_1 \right) - 1 ~, \qquad \lambda (q_1) \equiv \frac{H^2 q_1}{2} ~.
\end{equation}
The on-shell action is
\begin{equation}
\bar{S}_0 (q_1,\phi_1)  = \mp \frac{2}{H^2} \left[ (F-\lambda)^{3/2} \pm (F+\lambda)^{3/2} \right] ~.
\end{equation}
The saddle points \eqref{scalarsaddles} of the lapse integral are real when $F \geq \lambda$. A sufficient condition on $\phi_1$ for this to hold is $\phi_1^2 \geq 6 / H^2 q_1$. The corresponding solutions $\left( \bar{q}(\tau), \bar{\phi}(\tau) \right)$ in this regime involve large values of the scalar field where the potential is steep and the slow roll conditions are not satisfied. They all exhibit singularities at intermediate times, however, and thus do not specify valid, regular saddle points of the no-boundary wave function.\footnote{This can readily be seen from an explicit form of the saddle point histories $\left( \bar{q}(\tau), \bar{\phi}(\tau) \right)$, which are easily obtained from the solutions $\left( \bar{x}(\tau), \bar{y}(\tau) \right)$. See also \cite{Garay1990}, Section VI E.} It is well-known indeed (see e.g. \cite{HHH2008,Hartle:2007gi}) that the no-boundary wave function selects those Lorentzian histories that originate in relatively flat, inflationary patches of the scalar potential. This set of histories is associated with regular complex saddle points which probe the lower part of the potential where the slow roll conditions hold.

On the other hand if $F \leq - \lambda$ then the saddle points \eqref{scalarsaddles} are purely imaginary. This is the region of configuration space where the scale factor is small, and the exponent $i \bar{S}_0$ is real. Evaluating the wave function in this regime reveals it exhibits the familiar  exponentially growing behavior as a function of the scale factor \cite{Hawking1983}. 

We expect therefore that classical cosmological evolution will emerge as a prediction of the no-boundary wave function only in the regime $- \lambda < F < \lambda$ in which the scale factor is sufficiently large, $\lambda > 1/2$, and the scalar field is sufficiently small, $\phi_1 < \sqrt{3/2 \lambda}$.

In this classical domain of configuration space there are always two saddle points in the upper half complex $N$-plane and two in the lower half, just like in the de Sitter minisuperspace model. In the semiclassical limit, a Picard-Lefschetz analysis shows that the $\mathcal{C}=(-\infty,+\infty)^{\downarrow}$ contour should always be deformed to go through the saddles in the lower half plane, giving rise to a real wave function and an exponentially enhanced weighting to leading order in $\hbar$. The two relevant saddle points are
\begin{equation}
	N_{\pm} = \frac{1}{H^2} \left( \pm \sqrt{\lambda + F} - i \sqrt{\lambda - F} \right) ~,
\end{equation}
with on-shell action
\begin{equation}
	\bar{S}_{\pm} = \frac{2}{H^2} \left[ \mp (\lambda + F)^{3/2} - i (\lambda - F)^{3/2} \right] ~.
\end{equation}
After deforming the contour $\mathcal C$ in \eqref{2fieldintegral} to run over the appropriate Lefschetz thimbles, we use the saddle point approximation to evaluate \eqref{2fieldintegral}. To perform the Gaussian integrals in the neighbourhood of the saddle points, the identities $|N_\pm| = \sqrt{q_1} / H$ and
\begin{equation}
	S_0''(N_\pm) = \frac{\mp 6 i}{N_\pm} \sqrt{\lambda^2 - F^2}
\end{equation}
are useful. Furthermore, one has $\theta_+ = (\alpha_+ + \pi)/2, \theta_- = \alpha_- / 2$, where $\theta_\pm$ is the angle of the thimble with the positive real $N$-axis near the saddle point $N_\pm$, and $\alpha_\pm \equiv \arg (N_\pm)$. Also $\alpha_- = - \pi - \alpha_+$, and we denote $\alpha_+ \equiv \alpha$. Using this we obtain
\begin{equation} \label{NBcoshweight}
\Psi_{NB}^{\ } (q_1,\phi_1) = \sqrt{\frac{8 \pi H}{3 \hbar \, \sqrt{q_1}}} \left( \lambda^2 - F^2 \right)^{-1/4} e^{4 \pi^2 (\lambda - F)^{3/2} / \hbar H^2} \cos \left( \frac{4 \pi^2}{\hbar H^2} \left( \lambda + F \right)^{3/2} + \frac{\alpha}{2} \right) \left[ 1 + \mathcal{O}(\hbar) \right] ~.
\end{equation}

An approximate form of the wave function can be derived in the corner $\lambda \gg 1, \phi_1 \ll 1 / \sqrt{\lambda}$ of the classical regime where we have
\begin{align}
	\left( \lambda - F \right)^{3/2} &= 1 - \frac{H^2 q_1}{4} \phi_1^2 + \mathcal{O} \left( (\sqrt{\lambda} \phi_1)^4 \right) ~, \label{l-Fexpansion} \\
	\left( \lambda + F \right)^{3/2} &= \left( H^2 q_1 - 1 \right)^{3/2} \left[ 1 + \frac{H^2 q_1}{4 \left( H^2 q_1 - 1 \right)} \phi_1^2 + \mathcal{O}\left( \phi_1^4 \right) \right] ~, \\
	\left( \lambda^2 - F^2 \right)^{-1/4} &= \left( H^2 q_1 - 1 \right)^{-1/4} \left[ 1 + \frac{H^2 q_1}{24} \left( \frac{H^2 q_1 - 2}{H^2 q_1 - 1} \right) \phi_1^2 + \mathcal{O} \left( (\sqrt{\lambda} \phi_1)^4 \right) \right] ~, \label{prefactor} \\
	\alpha &= \beta \left[ 1 - \frac{H^2 q_1}{12 \beta} \sqrt{H^2 q_1 - 1} \, \phi_1^2 + \mathcal{O} \left( (\sqrt{\lambda} \phi_1)^4 \right) \right] ~, \\
	\beta &= - \tan^{-1} \left( \left( H^2 q_1 - 1 \right)^{-1/2} \right) ~.
\end{align}
Therefore we find
\begin{align} \label{res}
\Psi_{NB}^{\ } (q_1,\phi_1) &= \mathcal{P} ~ \exp \left[ \frac{4 \pi^2}{\hbar H^2} \left( 1 - \frac{H^2 q_1}{4} \phi_1^2 + \mathcal{O} \left( (\sqrt{\lambda} \phi_1)^4 \right) \right) \right] \times \notag \\ &\cos\left[ \frac{4\pi^2}{\hbar H^2} \left( H^2 q_1 - 1 \right)^{3/2} \left( 1 + \frac{H^2 q_1}{4 \left( H^2 q_1 - 1 \right)} \phi_1^2 + \mathcal{O}\left( \phi_1^4 \right) \right) + \frac{\alpha}{2} \right] \left[ 1 + \mathcal{O}(\hbar) \right] ~,
\end{align}
where an approximate form of the prefactor ${\cal P}$ can be determined from \eqref{prefactor}.

The wave function is real and describes in this classical $\lambda \gg 1, \phi_1 \ll 1 / \sqrt{\lambda}$ domain two identical copies of an ensemble of slow roll inflationary universes that are asymptotically de Sitter \cite{Hawking1983,HHH2008,Hartle:2007gi}. These classical ensembles can be viewed as the time-reversal of each other. For every classical history in the first ensemble, associated with an integral curve following from the $e^{+iS}$ factor\footnote{We emphasize that the classical Lorentzian histories predicted by the wave function are real and therefore distinct from the complex saddle point histories specifying its semiclassical approximation \cite{HHH2008}.} in \eqref{res}, its time-reversed is in the second ensemble associated with the $e^{-iS}$ factor. The individual histories have the same probabilities in both ensembles, with a lower probability for histories with more scalar field driven inflation. If one thinks of one set of histories as expanding, in the other set they are contracting. Individual histories in both ensembles are not connected classically but may be connected by quantum evolution mediated by the Wheeler-DeWitt equation \cite{Hartle:2015bna,Bramberger:2017cgf}.

Eq. \eqref{res} shows that small homogeneous perturbations around the de Sitter saddle points are suppressed. Naively extrapolating this to larger values of the scalar field would seem to suggest that the saddles in the upper half $N$-plane might become relevant for large perturbations $\phi_1 \sim 1/\sqrt{\lambda}$. However the general form of the wave function \eqref{NBcoshweight} differs drastically from the extrapolation of the perturbative result and shows this is not the case. The saddles in the upper half $N$-plane are exponentially suppressed in the entire classical domain of the wave function.\footnote{This conclusion holds for general scalar potentials \cite{HHH2008,Hartle:2007gi}.} More generally we note that solutions to the Wheeler-DeWitt equation are normalizable in the so-called induced inner product which essentially means the wave function must be an eigenstate of the Wheeler-DeWitt operator. This forbids wave functions from rapidly growing at the boundaries of superspace.\footnote{For a useful review of the induced inner product, see \cite{Embacher:1997zi} and references therein; for its implementation in quantum cosmology see \cite{Halliwell2009PRD}.}

Finally we turn to the wave function $\Psi$ for spatially varying scalar perturbations. For simplicity we consider just a single scalar perturbation mode around the de Sitter background in the ensemble. The action for this is given by\footnote{We employ the normalization convention $\int \dd^3 \Omega \sqrt{\Omega} \, Y_{\boldsymbol{k}} Y_{\boldsymbol{q}} = 2 \pi^2 \delta_{\boldsymbol{k},\boldsymbol{q}}$ for spherical harmonics on the three-sphere.}
\begin{equation} \label{perturbationaction}
S_l = \pi^2 \int_0^1 \dd \tau ~ N_\pm \left[\frac{q^2}{N_\pm^2}\dot{\phi}_l^2-l(l+2) \phi_l^2\right] ~,
\end{equation}
where $l>1$ labels the mode on the three-sphere of the perturbation under consideration, and $N_\pm$ are the two contributing background saddle points \eqref{dSMSPsaddles} in the lower half complex $N$-plane. We work to linear order, neglecting the backreaction of the scalar field on the metric. The regular saddle point solution for the perturbation mode is identical for the two background saddle point histories selected by the Picard-Lefschetz prescription and reads $\phi_l(\tau)=\phi_{l,1} f(\tau) / f(1)$ with \cite{FLT2}
\begin{align}
f(\tau) = \left(1 - \frac{i}{\tau H^2 N_\pm + i}\right)^{\frac{l}{2}}\left(1+\frac{i}{\tau H^2 N_\pm + i}\right)^{-\frac{l+2}{2}} \left(1+\frac{i(l+1)}{\tau H^2 N_\pm + i}\right) ~.
\end{align}
The on-shell perturbation action is
\begin{align}
i \bar{S}_l (q_1,\phi_{l,1})= \pi^2 \phi_{l,1}^2&\left[\mp i\frac{l(l+2)}{H}\sqrt{q_1}-\frac{l(l+1)(l+2)}{H^2} + \mathcal{O}\left(\frac{1}{\sqrt{q_1}}\right)\right] ~.
\end{align}
yielding
\begin{equation}
\Psi_{NB}^{\ } (q_1,\phi_{l,1})\sim e^{4\pi^2/\hbar H^2} e^{-\pi^2l(l+1)(l+2)\phi_{l,1}^2/\hbar H^2} \cos \left[\frac{4 \pi^2}{\hbar H^2}\left( \left( H^2 q_1 - 1 \right)^{3/2}+ \frac{l(l+2)}{4} \phi_{l,1}^2 \sqrt{H^2 q_1}\right)\right] ~.
\end{equation}
This result qualitatively generalizes to scalar and tensor perturbations around all classical backgrounds in both ensembles. Hence we recover the well-known Gaussian behaviour of the wave function of perturbations in the no-boundary state \cite{Halliwell:1984eu,Hartle:2010vi}.

The no-boundary condition of regularity on the perturbations implies that the Lorentzian perturbation histories exhibit growth in the direction of expansion in the classical backgrounds in both ensembles predicted by the wave function. This means that even if one were to connect both ensembles quantum mechanically, the physical arrows of time would reverse around the bounce \cite{Hawking:1993tu,HH2011}. This is in sharp contrast with the causality in ekpyrotic cosmology where the fluctuation arrow of time points in the same direction across the entire spacetime.

\section{Conclusion} \label{conclusionsec}
We have put forward a novel formulation of the no-boundary wave function that is based on a Lorentzian path integral. Using Picard-Lefschetz theory we have evaluated the Lorentzian path integral in the saddle point approximation in a homogeneous isotropic minisuperspace model consisting of gravity coupled to a positive cosmological constant and scalar field matter. With no-boundary conditions of regularity on geometry and field and with a contour for the lapse integral taken along the entire real axis, we recover the standard predictions of the semiclassical no-boundary wave function. Specifically, the resulting wave function is real and describes in its classical domain two copies of an ensemble of slow roll inflationary universes that are asymptotically de Sitter, with a Gaussian spectrum of small fluctuations.

Our results differ significantly from those of \cite{FLT1,FLT2} who also used Picard-Lefschetz theory to evaluate a Lorentzian minisuperspace path integral but with a contour for the lapse that runs over the positive real axis only. This choice of contour does not yield a solution of the Wheeler-DeWitt equation but rather a Green's function. More significantly, this contour choice is dominated by a saddle point that is different from those specifying the present model and which is, in particular, a ``wrong sign'' saddle point yielding fluctuation wave functions that imply fluctuations are not suppressed 
\cite{HarHal1990,Halliwell1988,Conti2015,FLT2}.

This choice fails to recover quantum field theory in curved space time \cite{HarHal1990}, and thus fails to provide a reasonable physical basis for a predictive framework for cosmology.

Note that the direction of the lapse integration is not directly related to the observed arrows of time such as those defined by the increase in entropy, the retardation of radiation, and the growth of fluctuations. As shown in \cite{Halliwell:1984eu} and as much subsequent work confirmed \cite{Hawking:1993tu,HH2011}, these physical arrows arise because the no-boundary wave function predicts that fluctuations are small when the universe was small. Histories of geometry are curves in the superspace of three-geometries. There is no physical notion of one three-geometry being `before' or `after' another. Reversing the sign of the lapse merely reverses the direction of parametrization of these curves without physical effect. None of these statements are at variance with the contribution of Teitelboim \cite{Teitelboim:1983fh} (the apparent motivation for the choice $N>0$ in \cite{FLT1,FLT2}), who suggested taking positive lapse purely by way of analogy to the familiar causal structure of Minkowski space time, but also acknowledged that this is a choice, not a necessity, thus leaving full freedom to explore the consequences of either half-infinite or infinite contours.

It would be interesting to explore whether the predictions of the Euclidean and the Lorentzian form of the no-boundary wave function differ beyond the semiclassical approximation. This would first require a more precise formulation of the wave function. A promising approach towards this is to use holographic techniques which might enable one to express the wave function of backgrounds and fluctuations in terms of the partition functions of a set of deformations of Euclidean CFTs defined on the future boundary. The reversal of the physical arrows of time in the histories predicted by the no-boundary wave function means it is conceivable that a single dual defined at future spacelike infinity indeed encodes all physical correlations.

\vskip .3cm
\noindent{\bf Acknowledgements:} We thank Jean-Luc Lehners and Neil Turok for correspondence. TH is supported in part by the National Science Foundation of Belgium (FWO) grant G092617N, by the C16/16/005 grant of the KULeuven and by the European Research Council grant no. ERC-2013-CoG 616732 HoloQosmos. JDD is supported by the National Science Foundation of Belgium (FWO) grant G.0.E52.14N Odysseus. OJ is supported by a James Arthur Fellowship.

\bibliographystyle{klebphys2}
\bibliography{references}
\end{document}